\documentclass{iopconfser}
\usepackage{siunitx}
\usepackage{graphicx}
\usepackage{subcaption}
\usepackage{cite}
\usepackage{amsmath}
\begin{document}

\title{Driver-delay chicanes for a multistage plasma-based accelerator facility}

\author{D Kalvik$^{1}$, E Adli$^{1}$, P Drobniak$^{1}$, F Peña$^{1,2}$ and C A Lindstrøm$^{1}$}

\affil{$^1$Department of Physics, University of Oslo, Oslo, Norway}

\affil{$^2$Ludwig Maximilian University of Munich, Munich, Germany}

\email{daniel.kalvik@fys.uio.no}

\begin{abstract}
The SPARTA project aims to design a medium-sized accelerator facility that facilitates new experiments in strong-field quantum electrodynamics using plasma-based accelerators. For this, we need several plasma stages and, therefore, several drivers. Drivers can be either an ultra-relativistic charged particle beam or a high-intensity laser beam. In case we use particle beams, we need a method of distributing these beams from a radio-frequency accelerator to the different plasma stages. A central part of this is a delay scheme that ensures temporal synchronization of the drivers. In this paper, we demonstrate how to achieve a \SI{2}{ns} delay in $\sim$\SI{12}{m}, while keeping the first-order beam parameters periodic.
\end{abstract}

\section{Introduction}
Plasma-wakefield acceleration is a method of accelerating a charged particle beam---here called the trailing bunch---by driving a wakefield in a plasma \cite{TajimaDawson, chen, Ruth:157249}. These wakefields are driven by a driver, which can be either a charged particle beam---called plasma wakefield acceleration (PWFA)---or an intense laser beam---called laser wakefield acceleration (LWFA)---and can achieve electric fields of up to \SI{100}{GV/m} \cite{leemans_gev_2006, leemans_multi-gev_2014, blumenfeld_energy_2007}. By leveraging these strong fields, a plasma-based particle accelerator could be made shorter and cheaper than a conventional accelerator. It works by having the driver deposit its energy into the plasma, which is then transferred to the trailing bunch via the electric fields. If the required energy is larger than what can be deposited by the driver, then we need several drivers and an equal number of plasma stages. This is referred to as staging. The SPARTA project~\cite{SPARTA} intends to design a medium-sized plasma-based accelerator facility using either PWFA or LWFA. Such a facility would facilitate new experiments within strong-field quantum electrodynamics (SFQED) \cite{SFQED, gonoskov_charged_2022}, simultaneously utilizing and demonstrating this new and cheaper technology.

\begin{figure}[htbp]
    \centering
    \includegraphics[width=1\linewidth]{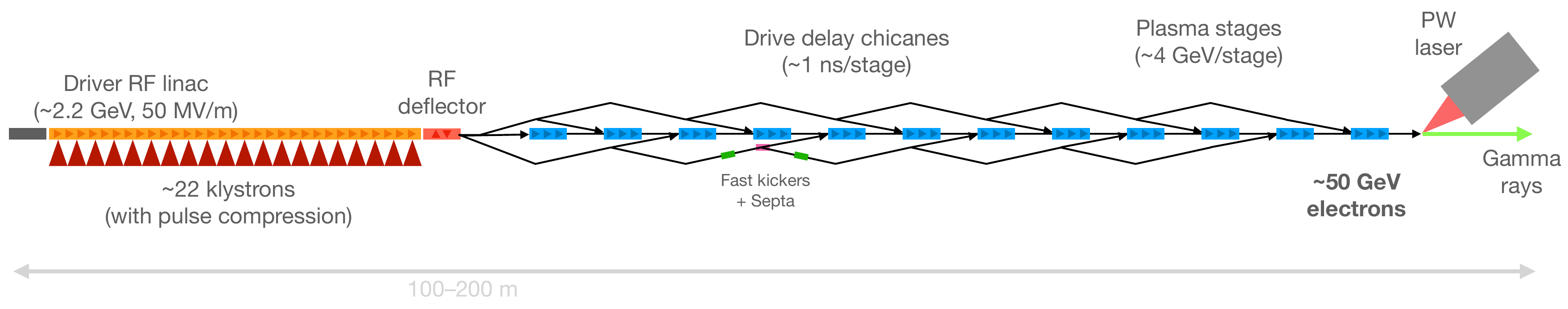}
    \caption{Sketch of a possible beam-driven SPARTA facility. The plasma stages are represented by the blue rectangles. The trailing bunch and the bunch train containing the drivers are accelerated by either a common, or two different, RF linacs (one common RF linac is shown here). An RF deflector (red rectangle) after the driver beamline (orange rectangle) separates every other driver into two different delay chicanes, allowing the fast kicker (green rectangle) to operate at a feasible bunch separation. The septum magnet is illustrated with the pink rectangle. At the downstream end, the accelerated trailing bunch would be used in beam--laser collisions with a petawatt-level (PW) laser.}
    \label{fig:SPARTA}
\end{figure}

If SPARTA is to use PWFA (instead of LWFA), we must ensure the properly synchronized delivery of the drivers into each plasma stage using delay chicanes. While different designs of delay chicanes have been considered previously \cite{adli_beam_2013, Pfingstner:IPAC2016-WEPMY010}, here we present a design specifically developed for the SPARTA project. We will therefore consider drivers of lower energy (\SI{2}{GeV}) than what is considered in the aforementioned references (\SI{25}{GeV}).
A sketch of the SPARTA facility is shown in Fig.~\ref{fig:SPARTA}, depicting two delay chicanes; one on each side of the accelerator beamline. Every other driver will be deflected into one of the two delay chicanes, by a radio-frequency (RF) deflector. This halves the bunch spacing in each delay chicane. The sketch assumes 12 plasma stages, each operating at a transformer ratio of 2 and boosting the energy of the trailing bunch by approximately \SI{4}{GeV}, with the final energy at approximately \SI{50}{GeV}.

\begin{figure}[htbp]
    \centering
    \includegraphics[width=0.8\linewidth]{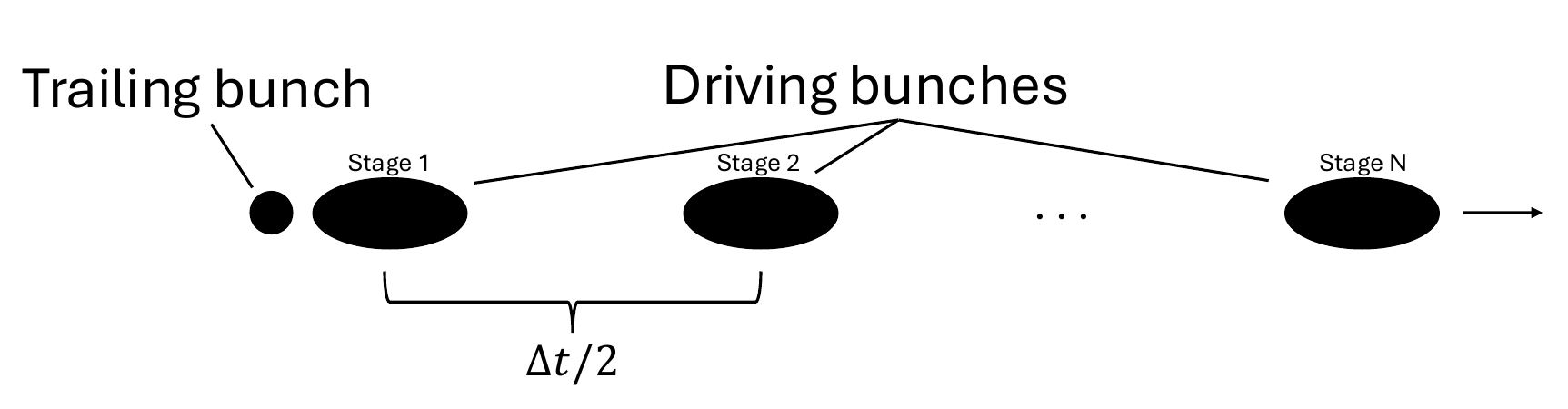}
    \caption{Sketch of the drive-train configuration with a trailing bunch coming out of the RF linac. The space between the trailing bunch and the last (most upstream driver) is synchronized for optimal acceleration in the plasma stage, while the distance between the drivers are all $\frac{\Delta t}{2}$. The direction of propagation is to the right, as indicated by the arrow.}
    \label{fig:bunches}
\end{figure}

The witness bunch and driving bunches are accelerated by an RF linear accelerator (linac)---not necessarily the same one---with a bunch spacing of $\frac{\Delta t}{2}$, as shown in Fig.~\ref{fig:bunches}. The factor of $\frac{1}{2}$ is used because we assume 2 delay chicanes, meaning we need each delay chicane to delay each driver by $\Delta t$ units of time. The reason we need this delay is that, for each stage, we need to adjust the temporal spacing between the trailing bunch and the driver. To explain this, we will consider every odd-numbered stage, i.e. stage number one, three, five, and so on. By design, the last (most upstream) driver is positioned at the correct temporal position relative to the trailing bunch, and will therefore be used in the first stage without the need for delay. However, if nothing is done by the third stage, the temporal distance between the trailing bunch and the driver will be $\Delta t$. By the fifth stage, the distance will be $2\Delta t$, and so on.
Longer delays are more challenging, but a certain minimum is required by extraction kickers (RF deflectors can, in principle, also be used, but are not considered here). The amount of space needed to achieve the required delay sets the minimum length of the combined plasma stage and interstage optics, effectively limiting the accelerating gradient of the facility. It is therefore imperative that a solution is found that allows for a long delay with a short amount of space.

\section{Theory}
Since the kickers limit the bunch-spacing of the drivers, we would like to use ultra-fast kickers. Such kickers are being developed for the PETRA IV project at DESY~\cite{Loisch:kickers, loisch:ipac2022-thpotk040}, and they allow a rise time of $\sim$\SI{2}{ns}. Although a shorter delay would be advantageous in designing a more compact delay chicane, the longer bunch spacing allows for reduced peak power in the klystrons, which helps to reduce the overall cost of the facility. Ultimately, the exact delay used will be subject to cost optimization and may therefore be adjusted.

\subsection{Kicks}
A kicker--septum arrangement is foreseen to be used to separate the driver prior to injection into the plasma stage. A septum magnet is a magnet that has a very thin separation between an area with a magnetic field and an area with no magnetic field~\cite{barnes2011injectionextractionmagnetssepta}. We can therefore use a kicker to give the last  bunch a different transverse position compared to the other bunches. We can then position the septum magnet so that the kicked bunch arrives at an area with no magnetic field, while the rest of the bunches continue to the next delay chicane.
To calculate the strength of the kicks we need for a septum magnet, we use the equation
\begin{equation}
    x_f = x_i + M_{12}\cdot x_i',
    \label{eq: M12}
\end{equation}
where $x_f$ is the final horizontal position (entering the septum), $x_i$ is the horizontal position of the beam at the time of the kick, $M_{12}$ is the positional dependence on the angle (a transfer matrix element), and $x_i'$ is the angle given by the kick. In the case that our lattice is not suitable for horizontal kickers, i.e., if $M_{12}$ is not sufficiently large, vertical kickers (in conjunction with Lambertson septa) will have to be considered. In this case, Eq.~\ref{eq: M12} is rewritten for the vertical plane, exchanging $x$ with $y$ and $M_{12}$ with $M_{34}$. According to Ref.~\cite{loisch:ipac2022-thpotk040}, for a \SI{2}{GeV} beam and a $\sim$\SI{0.4}{m} long kicker, $x_i'\simeq \SI{0.5}{mrad}$. Since a septum magnet requires at least \SI{2}{mm} separation \cite{barnes2011injectionextractionmagnetssepta}, we need an $M_{12} \geq \SI{4}{m}$. The actual values for our lattice are shown in Fig.~\ref{fig:M12/M34}.

\subsection{Delay}
\begin{figure}[htbp]
    \centering
    \includegraphics[width=0.6\linewidth]{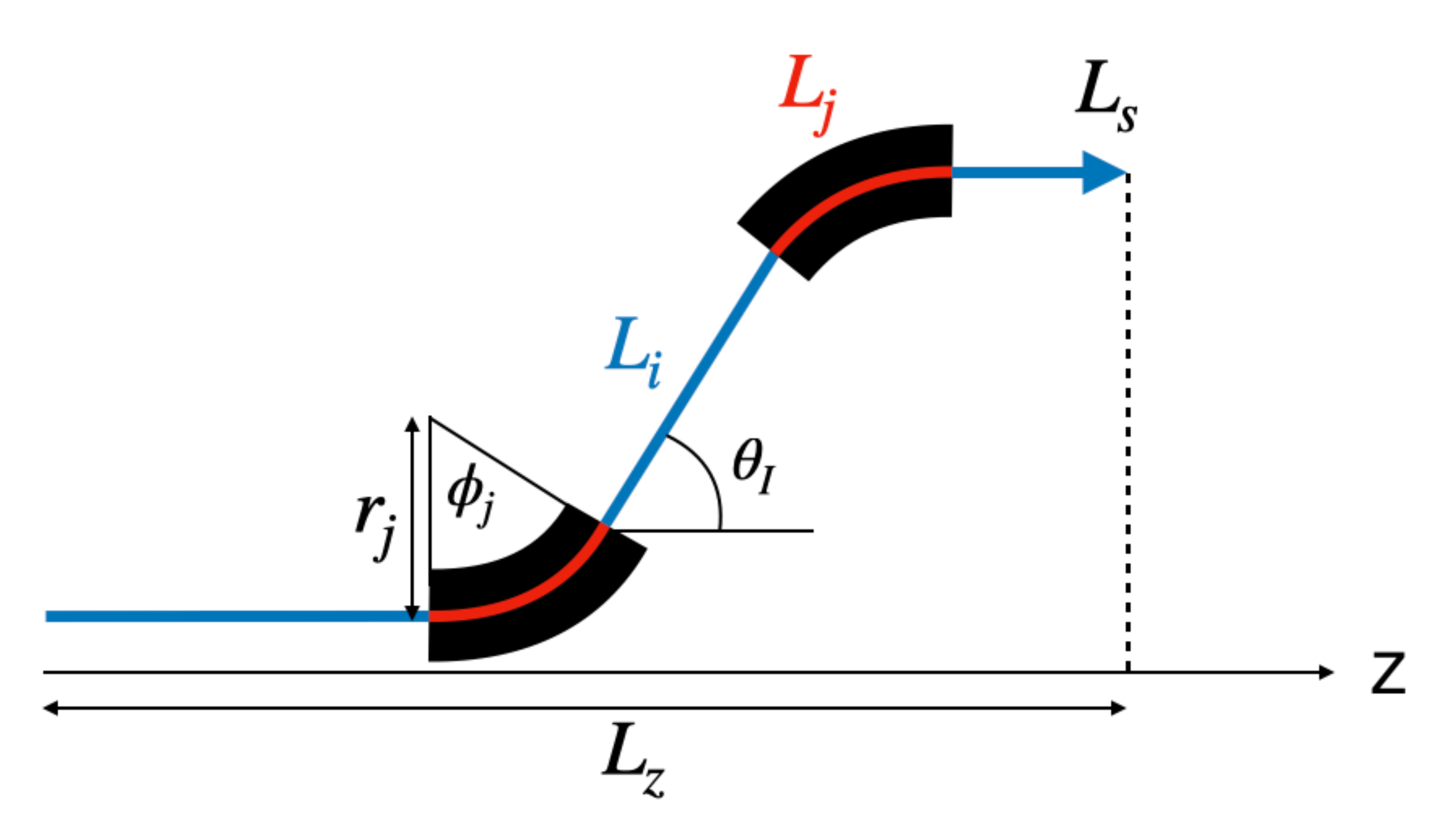}
    \caption{A sketch of dipoles and straight sections illustrating how the different variables of Eq.~\ref{eq:Delta-Ls} are defined.}
    \label{fig: delay equation params}
\end{figure}

Assuming ultra-relativistic beams, the delay is given by
\begin{equation}
    \Delta t = \frac{\Delta L}{c},
\end{equation}
where $c$ is the speed of light in vacuum, and $\Delta L$ is the length of the delay-chicane lattice minus the length of the accelerator beamline. $\Delta L$ can be calculated by taking the difference in the length of each individual beamline element and adding them all up. Although we are using three different elements---drift, quadrupole, and sector dipole---there are only two equations needed to calculate $\Delta L$. This is because, for a centered beam, quadrupoles only bend individual particles instead of the entire beam trajectory, and can therefore be regarded as drifts.
The two equations are then:
\begin{subequations}
\begin{gather}
    \Delta L_{\mathrm{drift}, i} = L_i(1-\cos(\theta_I)), \label{eq: L_straight}\\
    \Delta L_{\mathrm{bend}, j} = L_j - r_j(\sin(\phi_j + \theta_I) - \sin(\theta_I)). \label{eq: L_bend}
\end{gather}
\label{eq:Delta-Ls}
\end{subequations}
$\Delta L_{\mathrm{drift}, i}$ is the difference in length between the length of the element $L_i$, and its projection onto the longitudinal axis for straight section number~$i$. $\Delta L_{\mathrm{bend}, j}$ is the same, but applies to dipole number~$j$. $\theta_I$ is the angle of the beam coming into the element, $r_j$ is the bend-radius of dipole number~$j$, and $\phi_j$ is the bending angle of the same dipole. These parameters are shown in Fig.~\ref{fig: delay equation params}.
Using Eq.~\ref{eq: L_straight} and~\ref{eq: L_bend} and summing over all elements gives
\begin{equation}
    \Delta L = L_s - L_z= \sum_{i}L_i\left(1-\cos(\theta_I)\right) + \sum_j \left[L_j - r_j\left(\sin(\phi_j + \theta_I) - \sin(\theta_I)\right)\right],
    \label{eq: delta L}
\end{equation}
where $L_s$ is the length of the delay chicane, and $L_z$ is the length of the accelerator beamline.

\subsection{Periodicity}
To use the same kicker system and injection optics between every stage, we require the beam parameters to be periodic. In this paper, we consider the Twiss parameters, first-order horizontal dispersion, and the longitudinal dispersion ($R_{56}$). However, further research should include more detailed simulations investigating how average energy, energy spread, and emittance are affected by processes such as coherent and incoherent synchrotron radiation and second order dispersion.

\section{Lattice}
\begin{figure}[htbp]
    \centering
    \includegraphics[width=1\linewidth]{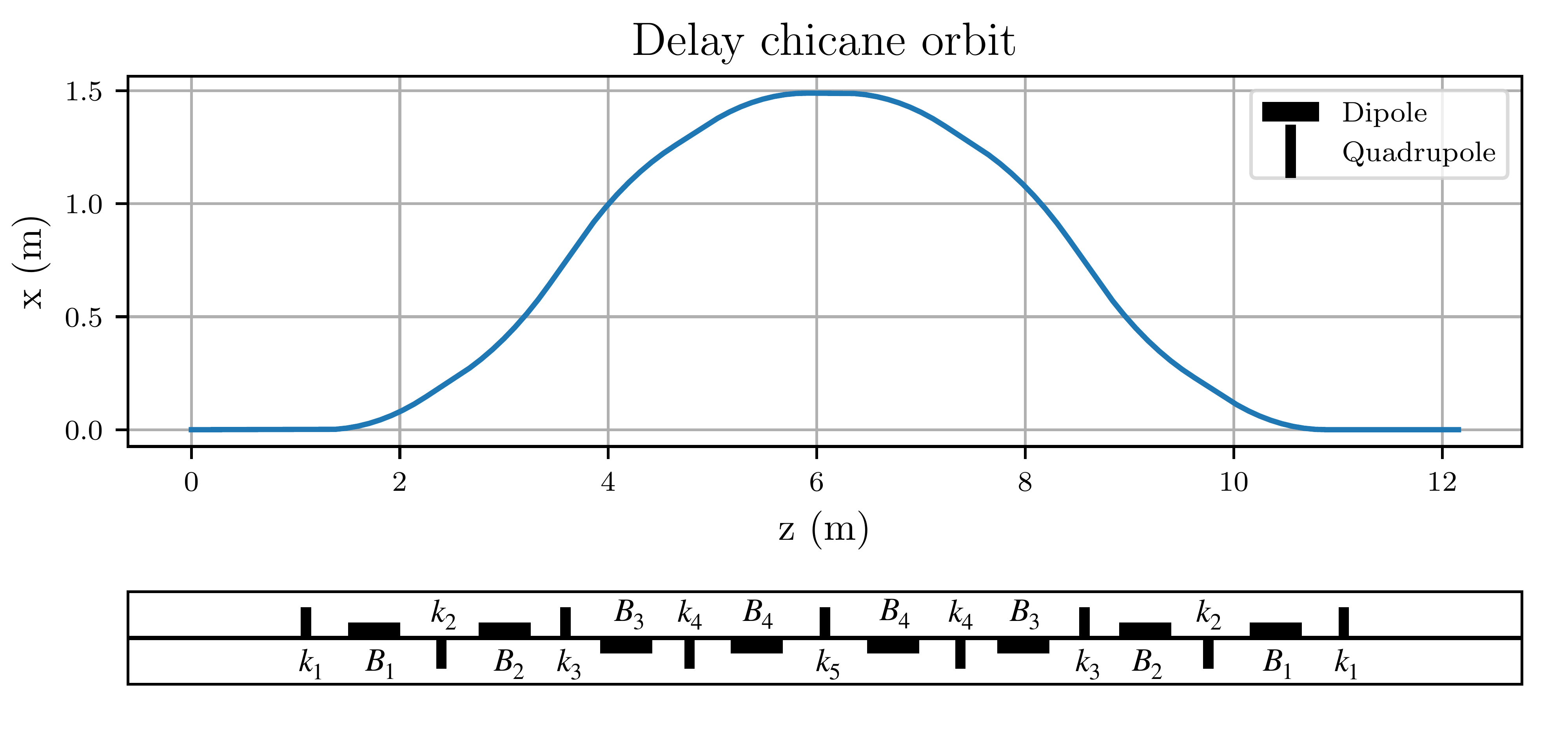}
    \caption{The horizontal orbit (top) of the delay chicane. The beamline lattice (bottom) follows the convention, where rectangles above the central axis have positive fields or are focusing in the horizontal plane, for dipoles and quadrupoles respectively. The achieved delay over a single chicane is \SI{2}{ns} for \SI{2}{GeV}, and all field strengths (or equivalent) are given in Table~\ref{tab: parameter values}. Note that the length of the indicated elements are not to scale to their actual lengths in the simulated lattice---only their mid-points are at the correct longitudinal location.}
    \label{fig: orbit lattice}
\end{figure}

The chosen lattice is an undulating chicane with 8 dipoles and 9 quadrupoles per two stages. There is a \SI{2}{m} gap every other stage, which is the designated space for the fast kicker. The lattice and orbit are shown in Fig.~\ref{fig: orbit lattice}.
The total periodic chicane is \SI{12.16}{m} projected onto the accelerator beamline, which allows for a plasma stage and interstage with a total length as small as \SI{6.08}{m}.

\subsection{Lattice parameters}
Table~\ref{tab: parameter values} shows the parameters of the delay-beamline elements and the energy of the driver. All dipoles are of the same length and strength. In addition, the drift spaces of the elements are: \SI{1}{m} to the first quadrupole, then \SI{7.5}{cm} between each element, all the way to the middle ($5^{\text{th}}$ quadrupole). The lattice is mirrored around the middle.
\begin{table}[htbp]
\caption{Lattice and beam parameters ordered by their appearance in the beamline. The lattice starts with an initial drift, followed by alternating quadrupoles and dipoles until the center quadrupole ($Q_5$), around which the lattice is mirror-symmetric. Geometric lengths are constant for each element type.}
\begin{center}
\begin{tabular}{|l|c|r|l|}
\hline
\textbf{Parameter Description} & \textbf{Symbol} & \textbf{Value} & \textbf{Unit} \\ \hline
Beam energy & $E$ & 2 & \SI{}{GeV} \\ \hline
\multicolumn{4}{|c|}{\textbf{Lengths}} \\ \hline
Initial drift length (to quadrupole 1) & $L_{d,0}$ & 1.0 & \SI{}{m} \\ \hline
Inter-element drift length & $L_{d}$ & 0.075 & \SI{}{m} \\ \hline
Quadrupole length & $L_Q$ & 20 & \SI{}{cm} \\ \hline
Dipole length & $L_B$ & 97 & \SI{}{cm} \\ \hline
\multicolumn{4}{|c|}{\textbf{Magnetic elements (up to symmetry point)}} \\ \hline
Quadrupole 1 strength & $k_1$ & 6.407 & \SI{}{\per\square\metre} \\ \hline
Dipole 1 magnetic field & $B_1$ & 2 & \SI{}{T} \\ \hline
Quadrupole 2 strength & $k_2$ & -6.526 & \SI{}{\per\square\metre} \\ \hline
Dipole 2 magnetic field & $B_2$ & 2 & \SI{}{T} \\ \hline
Quadrupole 3 strength & $k_3$ & 6.008 & \SI{}{\per\square\metre} \\ \hline
Dipole 3 magnetic field & $B_3$ & -2 & \SI{}{T} \\ \hline
Quadrupole 4 strength & $k_4$ & -7.958 & \SI{}{\per\square\metre} \\ \hline
Dipole 4 magnetic field & $B_4$ & -2 & \SI{}{T} \\ \hline
Quadrupole 5 strength (center) & $k_5$ & 15.601 & \SI{}{\per\square\metre} \\ \hline
\end{tabular}
\end{center}
\label{tab: parameter values}
\end{table}
We have considered permanent magnets of \SI{2}{T} in this paper; however, we will consider weaker magnets and, therefore, lower energy drivers or smaller delays in future research.

\subsection{Beam parameters}
\begin{figure}[htbp]
    \centering
    \includegraphics[width=1\linewidth]{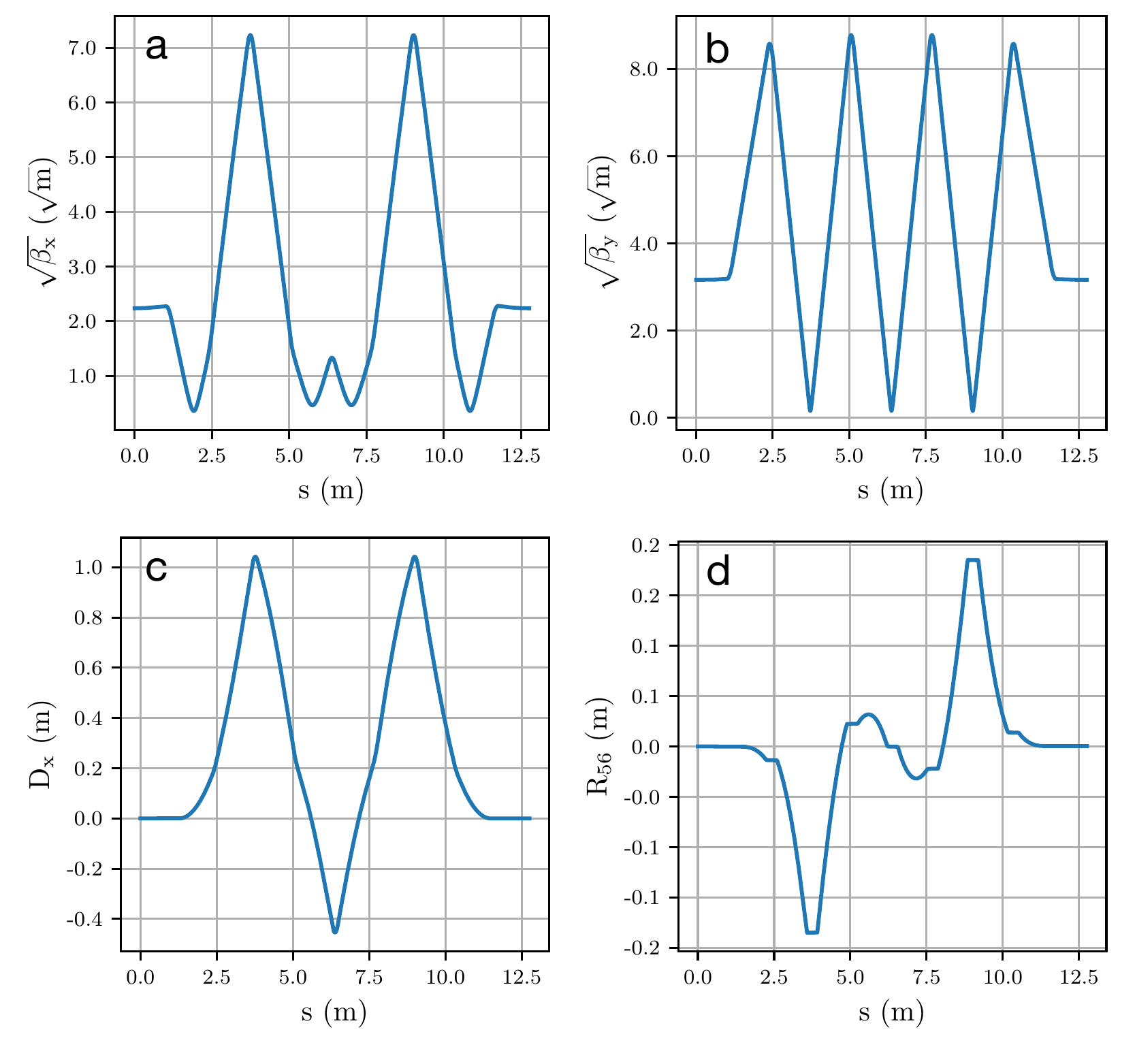}
    \caption{Evolution of (a) the horizontal $\beta$-function; (b) the vertical $\beta$-function; (c) the horizontal first-order dispersion; and (d) the longitudinal dispersion. All parameters shown are periodic, which is necessary for a repeating lattice.}
    \label{fig:twiss-dispersion-evolution}
\end{figure}

The periodicity of the beam parameters is achieved by optimizing each individual quadrupole, with the results shown in Fig.~\ref{fig:twiss-dispersion-evolution}.
The initial $\beta$-functions in the horizontal/vertical plane are \SI{5}{}/\SI{10}{m} and $\alpha_{x/y}=0$, while the first-order horizontal and longitudinal dispersions are both initially 0.
The figures show that all the considered beam parameters are periodic, which is required for the beam itself to be periodic. We note that the second-order horizontal dispersion is not considered here, which could be canceled by using sextupoles. However, this would require a slightly longer delay chicane.

\subsection{Positional dependency of kicks}
\label{sec: kicks}
Other important lattice parameters are the horizontal or vertical dependence on an initial angle in the corresponding plane. After the last  driver receives a kick from the kicker, we need to separate it from the rest of the train. This is accomplished by septum magnets, which, as previously mentioned, require at least \SI{2}{mm} separation.
\begin{figure}[t]
    \centering
    \includegraphics[width=1\linewidth]{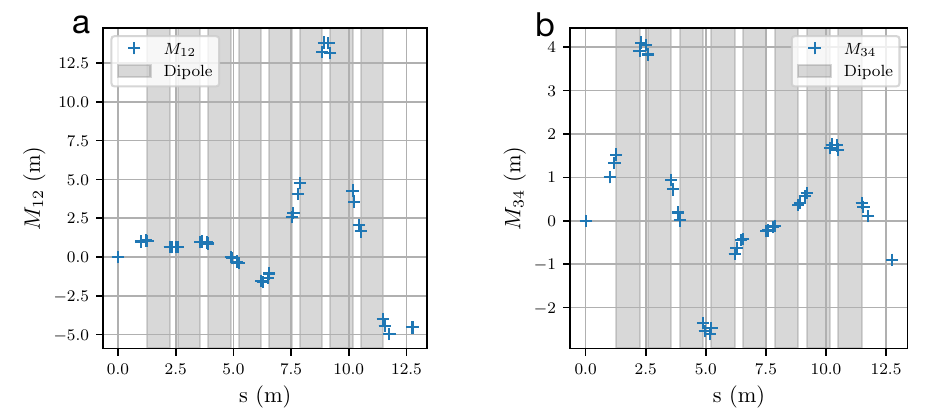}
    \caption{A plot of (a) $M_{12}$ and (b) $M_{34}$, as introduced in Eq.~\ref{eq: M12}. A comparison of (a) and (b) shows that a horizontal kicker is the best option to achieve a large separation for a septum magnet.}
    \label{fig:M12/M34}
\end{figure}
Using Eq.~\ref{eq: M12}, we showed that we need a minimum $M_{12}$ (or $M_{34}$ if kicked vertically) of \SI{4}{m}. According to Fig.~\ref{fig:M12/M34}, this is accomplished in both the horizontal and vertical planes, with the largest value obtained in the horizontal plane at the entrance to the $7^{\text{th}}$ dipole. Therefore, it is advantageous to use a horizontal kicker and to have the $7^{\text{th}}$ dipole be a septum magnet.

\section{$\beta$-mismatch tolerance}
\begin{figure}[h!]
    \centering
    \includegraphics[width=1\linewidth]{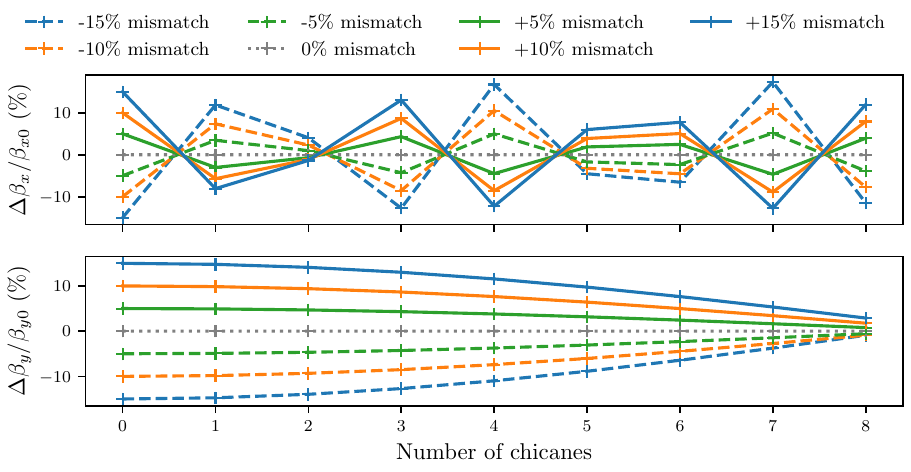}
    \caption{Outgoing $\beta$-mismatch in the horizontal plane (top) and the vertical plane (bottom) as a function of the number of delay chicanes. In the horizontal plane, the $\beta$-function is oscillating around the matched value, while the vertical plane shows a damping (or at least very slow oscillation) of the $\beta$-function.}
    \label{fig:mismatch}
\end{figure}

The stability of the beta-function is important to consider when we have a lattice that repeats several times. If a slight mismatch grows too quickly through the lattice, then the resulting beam will not have the required parameters for the injection optics and, subsequently, the plasma stage.
Figure~\ref{fig:mismatch} shows that the $\beta$-function for both planes is stable within a range of $\pm$\SI{15}{\%}, i.e., the resulting mismatch is smaller than or equal to the incoming mismatch. This means we can accept moderate mismatches through the delay chicane and hopefully through the injection system into the plasma---since the required $\beta$-values in plasma are not as strict for the drivers as for the trailing bunches.
In creating Fig.~\ref{fig:mismatch}, we have assumed the same distance in all stages and interstages; however, the length of the interstage must likely be scaled in accordance with the trailing-bunch energy. I.e., the interstage length and, therefore, the delay-chicane lattice will likely increase with the trailing-bunch energy. The scaling of this lattice with energy should, therefore, also be a point for future research.

\section{Conclusion}
In the case that the SPARTA facility were to use PWFA for acceleration in the plasma stages, the drivers need to be delayed as part of a larger distribution scheme. In this paper, we have shown a periodic first-order solution that provides the necessary \SI{2}{ns} delay over a length of about \SI{12}{m} for a \SI{2}{GeV} beam. The Twiss parameters, horizontal dispersion, and longitudinal dispersion are periodic. Specifically, we achieve zero horizontal and longitudinal dispersion in the straight section envisaged for a kicker. The lattice is also shown to be stable within a range of $\pm15\%$ mismatch of the $\beta$-functions, and $M_{12}$ is sufficiently large to allow the separation needed for a septum magnet. The presented solution, though specific to SPARTA, can easily be scaled to different energies, stage lengths, and delays. Future work in this regard will involve scaling the solution to be appropriate for the plasma-based collider concept HALHF \cite{foster_hybrid_2023, foster_proceedings_2025, adli:ipac2025-mocd1}.

\section*{Acknowledgments}
The authors wish to thank Jonas Bj\"orklund Svensson for his valuable input. This work is funded by the European Research Council (ERC Grant Agreement No.~101116161).

\clearpage

\bibliography{bib}

@inproceedings{Loisch:kickers,
    author = "Loisch, Gregor and Agapov, Ilya and Antipov, Sergey and Jebramcik, Marc and Keil, Joachim and Obier, Frank",
    title = "{Stripline Kickers for Injection Into PETRA IV}",
    booktitle = "{12th International Particle Accelerator Conference~}",
    doi = "10.18429/JACoW-IPAC2021-WEPAB113",
    month = "8",
    year = "2021"
}

@inproceedings{loisch:ipac2022-thpotk040,
  author       = {G. Loisch and V. Belokurov and F. Obier},
  title        = {{Few-Nanosecond Stripline Kickers for Top-Up Injection into PETRA IV}},
  booktitle    = {Proc. IPAC'22},
  pages        = {2858--2861},
  eid          = {THPOTK040},
  language     = {english},
  venue        = {Bangkok, Thailand},
  series       = {International Particle Accelerator Conference},
  number       = {13},
  publisher    = {JACoW Publishing, Geneva, Switzerland},
  month        = {07},
  year         = {2022},
  issn         = {2673-5490},
  isbn         = {978-3-95450-227-1},
  doi          = {10.18429/JACoW-IPAC2022-THPOTK040},
  url          = {https://jacow.org/ipac2022/papers/thpotk040.pdf},
}

@article{SFQED,
	title = {Experimental Signatures of the Quantum Nature of Radiation Reaction in the Field of an Ultraintense Laser},
	volume = {8},
	url = {https://link.aps.org/doi/10.1103/PhysRevX.8.031004},
	doi = {10.1103/PhysRevX.8.031004},
	pages = {031004},
	number = {3},
	journaltitle = {Physical Review X},
	shortjournal = {Phys. Rev. X},
	author = {Poder, K. and Tamburini, M. and Sarri, G. and Di Piazza, A. and Kuschel, S. and Baird, C.D. and Behm, K. and Bohlen, S. and Cole, J.M. and Corvan, D.J. and Duff, M. and Gerstmayr, E. and Keitel, C.H. and Krushelnick, K. and Mangles, S.P.D. and {McKenna}, P. and Murphy, C.D. and Najmudin, Z. and Ridgers, C.P. and Samarin, G.M. and Symes, D.R. and Thomas, A.G.R. and Warwick, J. and Zepf, M.},
	urldate = {2026-01-08},
	date = {2018-07-05},
	note = {Publisher: American Physical Society},
}

@InProceedings{Pfingstner:IPAC2016-WEPMY010,
  author       = {J. Pfingstner and E. Adli and C.A. Lindstrøm and E. Marín and D. Schulte},
  title        = {{C}onsiderations for a {D}rive {B}eam {S}cheme for a {P}lasma {W}akefield {L}inear {C}ollider},
  booktitle    = {Proc. of International Particle Accelerator Conference (IPAC'16),
                  Busan, Korea, May 8-13, 2016},
  pages        = {2565--2568},
  paper        = {WEPMY010},
  language     = {english},
  venue        = {Busan, Korea},
  series       = {International Particle Accelerator Conference},
  number       = {7},
  publisher    = {JACoW},
  address      = {Geneva, Switzerland},
  month        = {June},
  year         = {2016},
  isbn         = {978-3-95450-147-2},
  doi          = {doi:10.18429/JACoW-IPAC2016-WEPMY010},
  url          = {http://jacow.org/ipac2016/papers/wepmy010.pdf},
  note         = {doi:10.18429/JACoW-IPAC2016-WEPMY010},
}

@misc{adli_beam_2013,
	title = {A Beam Driven Plasma-Wakefield Linear Collider: From Higgs Factory to Multi-{TeV}},
	url = {http://arxiv.org/abs/1308.1145},
	doi = {10.48550/arXiv.1308.1145},
	shorttitle = {A Beam Driven Plasma-Wakefield Linear Collider},
	number = {{arXiv}:1308.1145},
	publisher = {{arXiv}},
	author = {Adli, Erik and Delahaye, Jean-Pierre and Gessner, Spencer J. and Hogan, Mark J. and Raubenheimer, Tor and An, Weiming and Joshi, Chan and Mori, Warren},
	urldate = {2026-01-08},
	date = {2013-09-29},
	langid = {english},
	eprinttype = {arxiv},
	eprint = {1308.1145 [physics]},
}

@inproceedings{SPARTA,
    author = {C. Lindstrøm and others},
    title = {The SPARTA project: toward a demonstrator facility for multistage plasma acceleration},
    booktitle = {Proc. IPAC'25},
    pages = {1615-1618},
    paper = {TUPS120},
    venue = {Taipei, Taiwan},
    series = {IPAC'25 - 16th International Particle Accelerator Conference},
    number = {16},
    publisher = {JACoW Publishing, Geneva, Switzerland},
    month = {06},
    year = {2025},
    issn = {2673-5490},
    isbn = {978-3-95450-248-6},
    doi = {10.18429/JACoW-IPAC2025-TUPS120},
    url = {https://indico.jacow.org/event/81/contributions/7824},
    language = {English},
    eventdate = {2025-06-01/2025-06-06},
}

@article{Ruth:157249,
      author        = "Ruth, Ronald D and Chao, A W and Morton, P L and Wilson, P
                       B",
      title         = "{A plasma wake field accelerator}",
      reportNumber  = "SLAC-PUB-3374",
      journal       = "Part. Accel.",
      volume        = "17",
      pages         = "171",
      year          = "1984",
}

@article{TajimaDawson,
  title = {Laser Electron Accelerator},
  author = {Tajima, T. and Dawson, J. M.},
  journal = {Phys. Rev. Lett.},
  volume = {43},
  issue = {4},
  pages = {267--270},
  numpages = {0},
  year = {1979},
  month = {07},
  publisher = {American Physical Society},
  doi = {10.1103/PhysRevLett.43.267},
  url = {https://link.aps.org/doi/10.1103/PhysRevLett.43.267}
}

@article{chen,
  title = {Acceleration of Electrons by the Interaction of a Bunched Electron Beam with a Plasma},
  author = {Chen, Pisin and Dawson, J. M. and Huff, Robert W. and Katsouleas, T.},
  journal = {Phys. Rev. Lett.},
  volume = {54},
  issue = {7},
  pages = {693--696},
  numpages = {0},
  year = {1985},
  month = {02},
  publisher = {American Physical Society},
  doi = {10.1103/PhysRevLett.54.693},
  url = {https://link.aps.org/doi/10.1103/PhysRevLett.54.693}
}

@misc{barnes2011injectionextractionmagnetssepta,
      title={Injection and extraction magnets: septa}, 
      author={M. J. Barnes and J. Borburgh and B. Goddard and M. Hourican},
      year={2011},
      eprint={1103.1062},
      archivePrefix={arXiv},
      primaryClass={physics.acc-ph},
      url={https://arxiv.org/abs/1103.1062}, 
}

@article{leemans_gev_2006,
	title = {{GeV} electron beams from a centimetre-scale accelerator},
	volume = {2},
	rights = {2006 Springer Nature Limited},
	issn = {1745-2481},
	url = {https://www.nature.com/articles/nphys418},
	doi = {10.1038/nphys418},
	pages = {696--699},
	number = {10},
	journaltitle = {Nature Physics},
	shortjournal = {Nature Phys},
	author = {Leemans, W. P. and Nagler, B. and Gonsalves, A. J. and Tóth, Cs and Nakamura, K. and Geddes, C. G. R. and Esarey, E. and Schroeder, C. B. and Hooker, S. M.},
	urldate = {2026-01-20},
	date = {2006-10},
	langid = {english},
	note = {Publisher: Nature Publishing Group},
}

@article{leemans_multi-gev_2014,
	title = {Multi-{GeV} Electron Beams from Capillary-Discharge-Guided Subpetawatt Laser Pulses in the Self-Trapping Regime},
	volume = {113},
	rights = {http://link.aps.org/licenses/aps-default-license},
	issn = {0031-9007, 1079-7114},
	url = {https://link.aps.org/doi/10.1103/PhysRevLett.113.245002},
	doi = {10.1103/PhysRevLett.113.245002},
	pages = {245002},
	number = {24},
	journaltitle = {Physical Review Letters},
	shortjournal = {Phys. Rev. Lett.},
	author = {Leemans, W.P. and Gonsalves, A.J. and Mao, H.-S. and Nakamura, K. and Benedetti, C. and Schroeder, C.B. and Tóth, Cs. and Daniels, J. and Mittelberger, D.E. and Bulanov, S.S. and Vay, J.-L. and Geddes, C.G.R. and Esarey, E.},
	urldate = {2026-01-20},
	date = {2014-12-08},
	langid = {english},
}

@article{blumenfeld_energy_2007,
	title = {Energy doubling of 42 {GeV} electrons in a metre-scale plasma wakefield accelerator},
	volume = {445},
	rights = {2006 Springer Nature Limited},
	issn = {1476-4687},
	url = {https://www.nature.com/articles/nature05538},
	doi = {10.1038/nature05538},
	pages = {741--744},
	number = {7129},
	journaltitle = {Nature},
	author = {Blumenfeld, Ian and Clayton, Christopher E. and Decker, Franz-Josef and Hogan, Mark J. and Huang, Chengkun and Ischebeck, Rasmus and Iverson, Richard and Joshi, Chandrashekhar and Katsouleas, Thomas and Kirby, Neil and Lu, Wei and Marsh, Kenneth A. and Mori, Warren B. and Muggli, Patric and Oz, Erdem and Siemann, Robert H. and Walz, Dieter and Zhou, Miaomiao},
	urldate = {2026-01-20},
	date = {2007-02},
	langid = {english},
	note = {Publisher: Nature Publishing Group},
}

@article{gonoskov_charged_2022,
	title = {Charged particle motion and radiation in strong electromagnetic fields},
	volume = {94},
	issn = {0034-6861, 1539-0756},
	url = {https://link.aps.org/doi/10.1103/RevModPhys.94.045001},
	doi = {10.1103/RevModPhys.94.045001},
	pages = {045001},
	number = {4},
	journaltitle = {Reviews of Modern Physics},
	shortjournal = {Rev. Mod. Phys.},
	author = {Gonoskov, A. and Blackburn, T. G. and Marklund, M. and Bulanov, S. S.},
	urldate = {2026-01-26},
	date = {2022-10-07},
	langid = {english},
}

@inproceedings{adli:ipac2025-mocd1,
    author = {E. Adli and others},
    title = {Updated baseline design for HALHF: the hybrid, asymmetric, linear Higgs factory},
    booktitle = {Proc. IPAC'25},
    pages = {53-56},
    paper = {MOCD1},
    venue = {Taipei, Taiwan},
    series = {IPAC'25 - 16th International Particle Accelerator Conference},
    number = {16},
    publisher = {JACoW Publishing, Geneva, Switzerland},
    month = {06},
    year = {2025},
    issn = {2673-5490},
    isbn = {978-3-95450-248-6},
    doi = {10.18429/JACoW-IPAC2025-MOCD1},
    url = {https://indico.jacow.org/event/81/contributions/7279},
    language = {English},
    eventdate = {2025-06-01/2025-06-06},
}

@article{foster_hybrid_2023,
	title = {A hybrid, asymmetric, linear Higgs factory based on plasma-wakefield and radio-frequency acceleration},
	volume = {25},
	url = {https://doi.org/10.1088/1367-2630/acf395},
	doi = {10.1088/1367-2630/acf395},
	pages = {093037},
	number = {9},
	journaltitle = {New Journal of Physics},
	publisher = {{IOP} Publishing},
	author = {Foster, B and D’Arcy, R and Lindstrøm, C A},
	date = {2023-09},
}

@article{foster_proceedings_2025,
	title = {Proceedings of the Erice workshop: A new baseline for the hybrid, asymmetric, linear Higgs factory {HALHF}},
	volume = {23},
	issn = {26660326},
	url = {https://linkinghub.elsevier.com/retrieve/pii/S2666032625000110},
	doi = {10.1016/j.physo.2025.100261},
	shorttitle = {Proceedings of the Erice workshop},
	pages = {100261},
	journaltitle = {Physics Open},
	shortjournal = {Physics Open},
	author = {Foster, Brian and Adli, Erik and Barklow, Timothy L. and Berggren, Mikael and Boogert, Stewart and Chen, Jian Bin Ben and D’Arcy, Richard and Drobniak, Pierre and Farrington, Sinead and Gessner, Spencer and Hogan, Mark J. and Kalvik, Daniel and Laudrain, Antoine and Lindstrøm, Carl A. and List, Benno and List, Jenny and Lu, Xueying and Pick, Gudrid Moortgat and Põder, Kristjan and Seryi, Andrei and Sjobak, Kyrre and Thévenet, Maxence and Walker, Nicholas J. and Wood, Jonathan},
	urldate = {2026-01-27},
	date = {2025-05},
	langid = {english},
}

\bibliographystyle{IEEEtran}

\end{document}